\documentclass[%
 reprint,
 amsmath,amssymb,
 aps,onecolumn
]{revtex4-2}

\usepackage{graphicx}
\usepackage{dcolumn}
\usepackage{bm}
\usepackage{graphicx}
\usepackage[utf8]{inputenc}

\usepackage{times}
\usepackage{color}
\usepackage{booktabs}
\usepackage{bm}
\usepackage{amsmath}
\usepackage{txfonts}
\usepackage{setspace}

\begin{document}

\preprint{APS/123-QED}

\title{Simultaneous magnetic and electric Purcell enhancement \\in a hybrid metal-dielectric nanostructure}

\author{Lingxiao Shan$^1$}
\author{Qi Liu$^{1,2}$}%
\author{Yun Ma$^1$}
\author{Yali Jia$^1$}
\author{Hai Lin$^1$}
\author{Guowei Lu$^{1,2,3,4}$}
\author{Qihuang Gong$^{1,2,3,4}$}
\author{Ying Gu$^{1,2,3,4}$}
\email{ygu@pku.edu.cn}
\affiliation{$^1$State Key Laboratory for Mesoscopic Physics, Department of Physics, Peking University, Beijing 100871, China\\
$^2$Frontiers Science Center for Nano-optoelectronics $\&$ Collaborative Innovation Center of Quantum Matter $\&$ Beijing Academy of Quantum Information Sciences, Peking University, Beijing 100871, China\\
$^3$Collaborative Innovation Center of Extreme Optics, Shanxi University, Taiyuan, Shanxi 030006, China\\
$^4$Peking University Yangtze Delta Institute of Optoelectronics, Nantong 226010, China\\
$*$ygu@pku.edu.cn}

\date{\today}

\begin{abstract}
Hybrid metal-dielectric structures, which combine the advantages of both metal and dielectric materials, support high-confined but low-loss magnetic and electric resonances under deliberate arrangements. However, their potential for enhancing magnetic emission has not been explored. Here, we study the simultaneous magnetic and electric Purcell enhancement supported by a hybrid structure consisting of a dielectric nanoring and a silver nanorod. 
Such a structure enables low Ohmic loss and highly-confined field under the mode hybridization of magnetic resonances on nanoring and electric resonances on nanorod in the optical communication band. So, the 60-fold magnetic Purcell enhancement and 45-fold electric Purcell enhancement can be achieved simultaneously with $>95\%$ of the radiation transmitted to far field. The position of emitter has a several-ten-nanometer tolerance for sufficiently large Purcell enhancement, which brings convenience to experimental fabrications. Moreover, an array formed by this hybrid nanostructure can further enhance the magnetic Purcell factors. The findings provide a possibility to selectively excite the magnetic and electric emission in integrated photon circuits. It may also facilitate brighter magnetic emission sources and light-emitting metasurfaces in a simpler arrangement. 

\end{abstract}

\maketitle
\section{\label{sec:level1}Introduction}
Nowadays, the magnetic Purcell effect \cite{purcell_spontaneous_1946} is receiving more attention in photonic applications such as optical antennas \cite{albella_low-loss_2013} and metamaterials \cite{hussain_enhancing_2015}, which enables further employment and characterizing of magnetic emission in nanophotonic structures \cite{aigouy_mapping_2014}. 
In common systems, the strength of magnetic dipole transitions is usually weaker than electric dipole transitions by several orders \cite{baranov_modifying_2017}. But some emitters as lanthanide ions \cite{carnall_spectral_1968, ofelt_intensities_1962, judd_optical_1962} and quantum dots \cite{zurita-sanchez_multipolar_2002} have considerable magnetic dipole magnitudes \cite{brewer_coherent_2017} in the optical range, which can be used as magnetic quantum emitters (MQEs). To achieve strong magnetic Purcell effects, nanostructures supporting strong magnetic resonances are also necessary. Metallic split-ring resonators \cite{hein_tailoring_2013, fang_coherent_2016}, dielectric nanospheres \cite{rolly_promoting_2012, liu_control_2014, chigrin_emission_2016, sugimoto_magnetic_2021}, dielectric nanodisks \cite{staude_tailoring_2013, feng_all-dielectric_2016, wang_broadband_2020}, dimer antennas \cite{albella_low-loss_2013, bakker_magnetic_2015, zywietz_electromagnetic_2015}, diabolo antennas \cite{grosjean_diabolo_2011, mivelle_strong_2015}, nanohole \& nanoparticle arrays \cite{hussain_enhancing_2015, cui_enhancement_2016, li_asymmetric_2021, jia_surface_2021} and wire magnetic metamaterials \cite{slobozhanyuk_magnetic_2014, mirmoosa_magnetic_2016} have been investigated for providing strong magnetic resonances with a high magnetic density of states (MLDOS).

The nanostructures mentioned above are made of only dielectric or metal, which have intrinsic advantages and disadvantages: dielectric nanostructures provide low intrinsic loss, high radiation directivity but weaker field enhancement, while metal ones have ultrastrong field confinement but substantial Ohmic loss. Nevertheless, hybrid nanostructures can balance low loss (higher quality factors) with high confinement (smaller mode volumes) between two kinds of structures under deliberate design of hybrid resonances. It has been recently reported that multilayer nanodisks \cite{wang_hybrid_2013, yang_low-loss_2017, yang_greatly_2019, sun_metal-dielectric_2020}, hybrid nanodisk-nanoparticle structures \cite{rusak_hybrid_2014, gili_metaldielectric_2018, sun_enhanced_2019, yang_strong_2020}, dimer antennas \cite{sun_metaldielectric_2017, karamlou_metal-dielectric_2018, sun_highly_2020}, Yagi-Uda antennas \cite{ho_highly_2018}, spoof nanodisks \cite{wu_strong_2019}, nanoparticle-cavity systems \cite{schmidt_hybrid_2012, zhu_highly_2018, zhang_chiral_2019, zhang_hybrid_2020, doeleman_observation_2020, shan_generation_2022, hu_strong_2021, barreda_hybrid_2022}, metal-semiconductor nanowires \cite{lian_efficient_2015, shan_large_2020} and hybrid gratings \cite{sarkar_hybridized_2019} are utilized as hybrid nanophotonic structures. These nanostructures have exhibited their potentials in nanolasers \cite{oulton_plasmon_2009, lu_plasmonic_2012, li_plasmonic_2019}, strong light-matter interaction \cite{zhang_chiral_2019, shan_generation_2022}, radiation directivity control \cite{wang_janus_2015, xu_dual-frequency_2019}, nonlinear effects \cite{shibanuma_efficient_2017, shi_efficient_2019} and sensing \cite{ray_hybrid_2020}. But these hybrid structures are mostly employed in producing considerable electric Purcell enhancement \cite{qian_spontaneous_2021}. The magnetic Purcell enhancement, though important, has been rarely studied. Their potentials in enhancing magnetic Purcell effect needs further investigation via combining the advantages of dielectric and metal materials.

Here, we propose a hybrid structure with a dielectric nanoring and a metal nanorod to efficiently enhance the emission from both magnetic and electric dipole transitions in the optical communication band. Strong magnetic and electric optical resonances are simultaneously supported in our hybrid structure. Compared to the single nanoring or nanorod, it can yield the higher magnetic ($\sim60$) and electric Purcell factors ($\sim 45$) resulting from larger magnetic and electric density of states (MLDOS \& ELDOS). 
Our structure also exhibits high radiation efficiency ($>95\%$) and may benefit far-field radiation manipulations. We find that the positions of magnetic (MQE) or electric quantum emitters (EQE) can range in a large region, which relaxes the restriction of experimental fabrications. Additionally, when hybrid structures are arranged as an array, MQE can yield a larger Purcell enhancement because of increasing LDOS. The structure provides an option of photon sources on integrated circuits with selective excitations of magnetic and electric emission \cite{li_plasmon-assisted_2018}. It can also boost novel applications based on strong magnetic light-matter interactions such as light-emitting metasurfaces and optical antennas for magnetic emission.

\section{\label{sec:level1}Model setup}
The resonances in our hybrid structure originate from the hybridization of a silicon nanoring and a silver nanorod. For the single nanoring or nanorod, the magnetic dipole (MD) resonance of nanoring \cite{feng_all-dielectric_2016} and the electric quadrupole (EQ) of nanorod \cite{curto_unidirectional_2010} are located at 1360 nm. These resonances enable very strong magnetic or electric resonances so that the emitter can yield very high Purcell enhancement around nanoring or nanorod. Under the mode hybridization, the magnetic-like and electric-like resonances appear at 1305 nm and 1430 nm, which demonstrate high magnetic and electric density of states (MLDOS \& ELDOS). Besides, Ohmic loss can also be suppressed under the mode hybridization. Thus, higher magnetic and electric Purcell enhancements can be simultaneously achieved with a high radiation efficiency. Particularly, the magnetic Purcell factor approaches 60 and the radiation efficiency keeps steady above 95\%. The large MLDOS are distributed in the region around nanoring so that the requirement of emitter’s position can be relaxed in experimental fabrications. Moreover, EQE can feel a 40-fold Purcell enhancement in the gap between nanoring and nanorod, and the emission enhancement can reach a higher level by narrowing the gap.

\begin{figure}
	\center
	\includegraphics[width=0.9\textwidth]{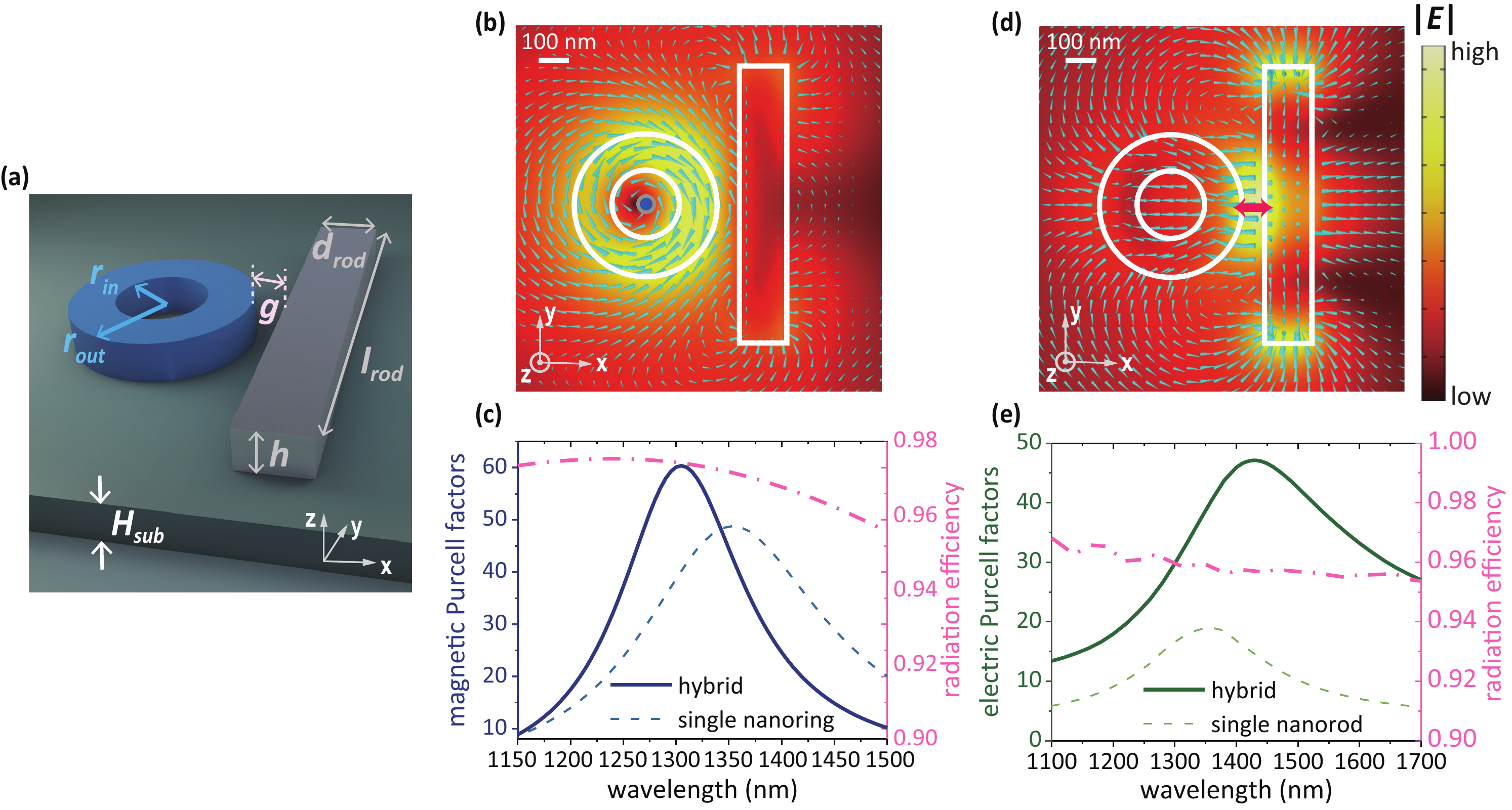}
	\caption{(a) Schematic of our hybrid structure consisting of a silicon nanoring ($n_{\rm ring}=3.7$), a silver nanorod and a silica substrate.
	Structural parameters are ticked out in the illustration. 
	(b) Field profile of the magnetic-like resonance excited by MQE (ticked out by a blue dot at the center of nanoring). (c) Purcell factor (solid line) and radiation efficiency (pink dash line) under the excitation of MQE. The peak of magnetic Purcell enhancement appears at 1305 nm. The hybrid structure obtains higher peak Purcell factors compared to a single nanoring excited by MQE, which is illustrated by a blue dash line. The radiation efficiency keeps above 95\%. (d) Field profile of the electric-like resonance excited by EQE (ticked out by a dark red arrow in the middle of gap). (e) Purcell factor (solid line) and radiation efficiency (pink dash line) under the excitation of EQE. The peak of electric Purcell enhancement appears at 1430 nm. The higher Purcell enhancement can be achieved in the hybrid structure than in the single nanorod (illustrated by a brown dash line). The $>90\%$ radiation efficiency can also be achieved.
	Cyan arrows depict electric field around the hybrid structure in (b, d).
	}
	\label{fig:Res}
\end{figure}

The hybrid structure consists of a silicon nanoring and a silver nanorod on a silica substrate as shown in Fig. \ref{fig:Res}(a). The two components are separated by a gap with $g=70 \,\rm nm$ (10-200 nm is suitable for high Purcell enhancement). An emitter is set 55 nm above substrate (ranging from 0 to 150 nm in the discussion part). The entire system is exposed in the air. The outer radius of silicon nanoring is $r_{\rm out}=225 \,\rm nm$ and the inner radius is $r_{\rm in}=110 \,\rm nm$ with its refractive index $n=3.7$. The size of silver nanorod is $l_{\rm rod}=870 \,\rm nm$ and $d_{\rm rod}=150 \,\rm nm$ with its permittivity from John and Christy in 1972 \cite{johnson_optical_1972}. The height of two components is $h=110 \,\rm nm$. The height of silica substrate is set as $H_{\rm sub}=100 \,\rm nm$ with its refractive index $n=1.48$ (ranging from 1 to 3 in the discussion part).

The electromagnetic simulation is conducted in the commercial COMSOL software using finite element analysis. The size of simulation module is set as $2700 \,\rm nm\times 2700 \,\rm nm\times 2100 \,\rm nm$. A 150-nm perfectly matched layer is used at the module boundary to simulate an infinite space. The 100-nm substrate is located in the middle of module with two 1-$\rm \mu m$ air layers locating in the top and bottom. The air and substrate are equipped with meshes of 70 nm and 50 nm, respectively. The nanoring and nanorod have a finer mesh of 20 nm. To calculate the total power, a 4-nm sphere surrounding the emitter is applied with a 2-nm mesh to obtain the energy flux integration $P_{\rm tot}=\oiint\vec{S}\cdot d\vec{A}$. The power of far-field radiation is calculated by the power flux integration on outer boundaries surrounding the entire module.

\section{\label{sec:level1}Magnetic and electric Purcell enhancement in the hybrid structure}
\subsection{\label{sec:level2}Resonances in the hybrid structure}

Our hybrid metal-dielectric structure supports the hybrid magnetic-like and electric-like resonances stemming from MD and EQ resonances in a single silicon nanoring and silver nanorod. Here, we set the resonances of individual nanoring or nanorod at 1360 nm (Structural parameters are listed in model setup part above, and more details are put in Appendices). For the silicon nanoring, MD resonance dominates under this wavelength. The previous research has shown that the nanoring can provide high Purcell enhancement with MQEs located in the center \cite{feng_all-dielectric_2016}. For a silver nanorod, only electric resonances are supported in rod geometries. The rod length is chosen as 870 nm where EQ resonance dominates. When the rod length increases, higher-order electric resonances appear successively. EQEs beside can yield strong Purcell enhancement \cite{ curto_unidirectional_2010}. 
Superior to the single nanoring and nanorod, our proposed structure enables hybrid resonances with a further increase of LDOS and low Ohmic loss.

When the two components are put together, two hybrid resonances move apart from original MD and EQ resonances. 
The first is the magnetic-like resonance located at 1305 nm [Fig. \ref{fig:Res}(c)]. It is excited by MQE at the center of nanoring and 55 nm above substrate, whose polarization is perpendicular to substrate. As shown in Fig. \ref{fig:Res}(b), the electric field around structure indicates that the resonance is a combination of MD in nanoring and ED in nanorod. The magnetic-like resonance manifests strong field confined inside and around nanoring but the nanorod has weaker excitation. The nanoring contributes a large fraction of MD resonance thus there is a high MLDOS around nanoring enabling a high magnetic Purcell enhancement. Additionally, the nanorod also contributes to the MLDOS enhancement which can be seen in a larger Purcell enhancement compared to the single nanoring. The second is the electric-like resonance located in 1430 nm, which is excited by EQE in the gap with in-plane polarization normal to nanorod [Fig. \ref{fig:Res}(d)]. Its field profile indicates a pattern of EQ in the nanorod, while the nanoring has a weaker excitation. Because of the high field confinement from gap surface plasmons, the electric-like resonance can support the strong electric Purcell enhancement, which surpasses the emission enhancement with a single nanorod [Fig. \ref{fig:Res}(e)].




\subsection{\label{sec:level2}Magnetic and electric Purcell enhancement}
Then we investigate the magnetic or electric Purcell effect when MQE or EQE is situated at the different points of hybrid nanostructure. For the magnetic Purcell effect, MQE is located in the hollow region of dielectric nanoring and its orientation is normal to substrate. Such a setup can maximally excite the magnetic-like resonance for the largest Purcell enhancement \cite{feng_all-dielectric_2016, sun_enhanced_2019}, which surpasses the Purcell factor obtained in a single nanoring. Compared with relevant work to enhance magnetic Purcell effects, the level of magnetic Purcell factor is superior or close to dielectric nanospheres \cite{sugimoto_magnetic_2021}, metal diabolo antennas \cite{mivelle_strong_2015} and metal nanodisk arrays \cite{cui_enhancement_2016}. And the distance (55 nm) between emitter and structure are larger than these previous research under similar Purcell enhancement, which brings convenience to nano-assembling. As shown in Fig. \ref{fig:Res}(c), the magnetic Purcell factor can reach a peak of 59 at 1305 nm [Fig. \ref{fig:Res}(c)] when MQE is located 55 nm above substrate. Compared with the single nanoring excited under the same condition, the peak Purcell enhancement is higher in the hybrid structure, and its linewidth is narrower. Moreover, the Purcell factor can be further improved with the decrease of gap width $g$, which will be discussed in the following section. To characterize the radiation property of hybrid structure, we adopt a radiation efficiency $\eta=P_{\rm rad}/P_{\rm tot}$ by measuring the portion of radiated power to far field. As shown in the purple line in Fig. \ref{fig:Res}(c), $>90\%$ of the emission from emitter can be transmitted to the far field region.

For the electric Purcell effect, EQE is located in the middle of gap and its dipole orientation is normal to nanorod, which corresponds to the hotspot of EQ resonance [Fig. \ref{fig:Res}(d)]. The two structures constitute a gap plasmon with large ELDOS \cite{hugall_plasmonic_2018}, so the Purcell enhancement can be further enhanced compared with a single nanorod. When the gap decreases, the electric Purcell factor displays a dramatic rise which is exhibited in the following part. The peak of electric Purcell factor is 46 at 1430 nm. It is worth mentioning that the radiation efficiency keeps at a very high level above 95\% [Fig. \ref{fig:Res}(e)], which is useful for far-field light manipulations. In our hybrid structure, MQE and EQE excite highly-confined resonances with low absorption, which results in the simultaneous strong magnetic and electric Purcell enhancement along with high radiation efficiency. Such advantages of the hybrid structure may advance efficient nanoscale photon sources in a simpler arrangement.

\subsection{\label{sec:level2}Influence of structural and material parameters on Purcell factors}
In the following, we discuss the influence for magnetic and electric Purcell enhancement under the variations of structural and material parameters. Here, the gap width $g$ modulates Purcell enhancement in different ways by altering the magnetic and electric-like resonances. To achieve high magnetic and electric Purcell enhancement simultaneously, $g$ should be set around 70 nm. The substrate refractive index $n$ influences the coupling between nanostructure and substrate. The Purcell factors will experience a slight transition with a red shift of resonance when $n$ increases. Moreover, a smaller inner radius $r_{\rm in}$ of nanoring can effectively enhance magnetic Purcell effect, and electric Purcell effect has a slight change under the increasing $r_{\rm in}$. These parameters provide various freedoms to modulate Purcell effect and resonance wavelengths in our hybrid structure. These freedoms can provide more possibilities in controlling photon sources on integrated circuits. 

\begin{figure}
	\center
	\includegraphics[width=0.85\textwidth]{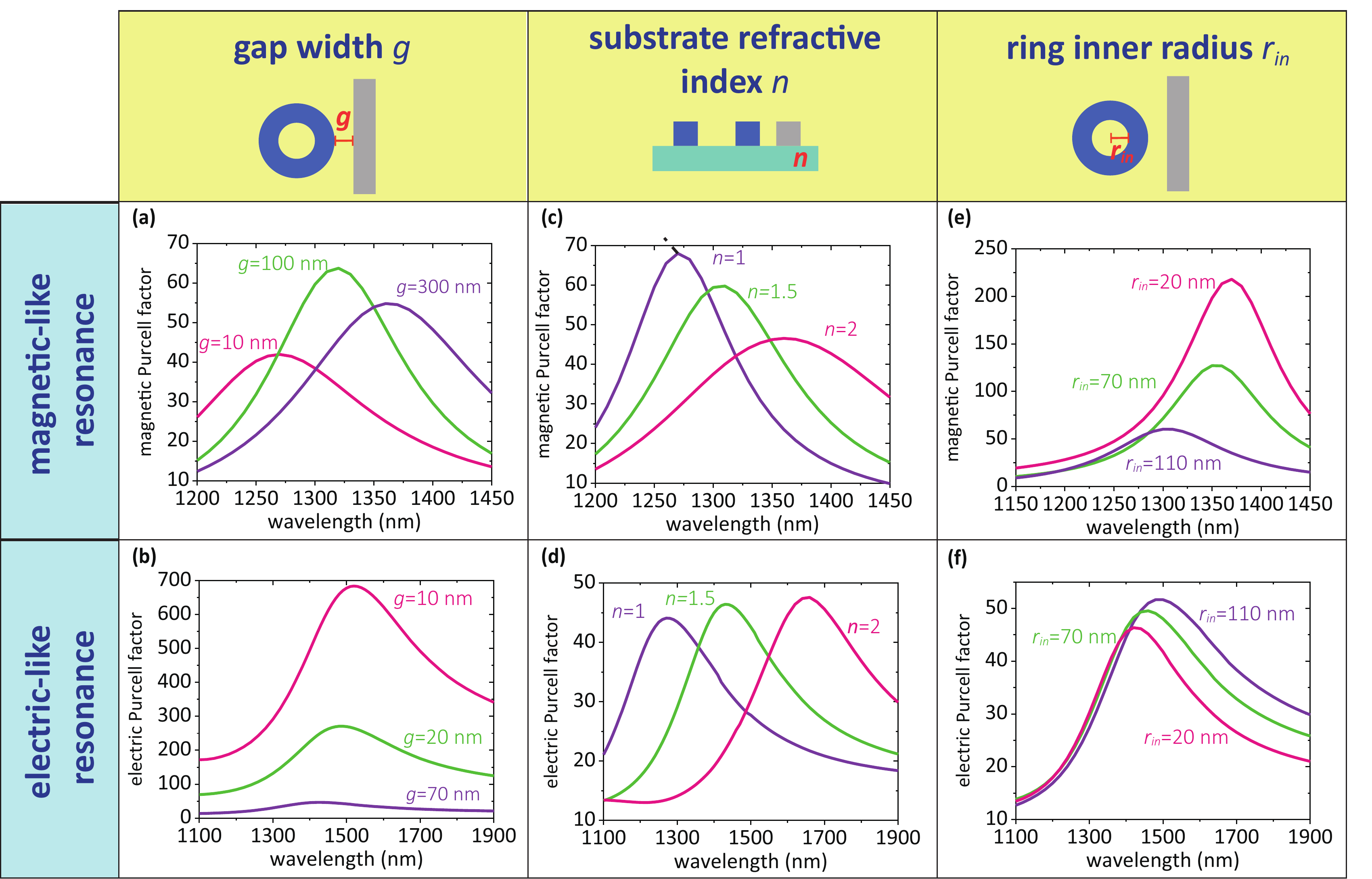}
	\caption{ Variations of the magnetic and electric Purcell enhancement under some structural parameters in the magnetic(electric)-like resonance. Magnetic and electric Purcell factors as a function of (a, b) gap width $g$; (c, d) substrate refractive index $n$; (e, f) inner radius of nanoring $r_{\rm in}$. 
	}
	\label{fig:Str}
\end{figure}

When the gap width $g$ varies, LDOS in two hybrid resonances are modified. The optimal choice of $g$ is around 70 nm for both high magnetic and electric Purcell enhancements. For the magnetic-like resonance [Fig. \ref{fig:Str}(a)], the magnetic Purcell factors are lower under a very narrow gap below 50 nm. This phenomenon results from the stronger mode hybridization between nanoring and nanorod so that MLDOS decreases but ELDOS occupies a larger fraction. The maximum of magnetic Purcell enhancement appears as 64 when $g$ is 100 nm. As the gap keeps increasing, MLDOS reduces to the level of single nanoring and the resonance wavelength has a red shift. Thus, the gap width should be moderate for the largest MLDOS in the hybrid structure, which is suitable for nanoscale assembling. While, for the electric-like resonance [Fig. \ref{fig:Str}(b)], the narrower gap brings much higher Purcell enhancement owing to the ultrahigh ELDOS in gap plasmon systems \cite{hugall_plasmonic_2018}. And the Purcell factor has a sharp increase with a red shift of the resonance wavelength when $g$ shrinks below 20 nm. Particularly, the Purcell factor approaches 680 at a wavelength of 1500 nm when $g=10 \, \rm nm$, which stays at the same level as that in all-metal two-layer diabolo antennas \cite{mivelle_strong_2015}.

The refractive index $n$ of substrate also has a significant influence on Purcell enhancement. When $n$ increases, the magnetic-like resonance experiences a red shift and the peak values of Purcell enhancement have a decline [Fig. \ref{fig:Str}(c)]. The maximal magnetic Purcell enhancement drops from 68 with $n=1$ to 40 with $n=2.5$. This is because the approaching refractive index between nanoring and substrate inducing the field extension into substrate. Such an extension changes confined field around nanostructures and then pushes these resonances to longer wavelengths. While, for the electric-like resonance, Purcell enhancement has only a slight rise with the increasing $n$. But the range of red shift of resonances reaches about 400 nm when $n$ ranges from 1 to 2 [Fig. \ref{fig:Str}(d)]. The substantial red shift can be attributed to the stronger hybridization of nanorod and substrate \cite{van_de_groep_designing_2013, spinelli_controlling_2011}.

The inner radius $r_{\rm in}$ of nanoring can strongly modulate MLDOS and enhance magnetic Purcell effect in the magnetic-like resonance, which was investigated in the single dielectric nanoring \cite{feng_all-dielectric_2016}. The phenomenon also works in the hybrid structure [Fig. \ref{fig:Str}(e)]. When $r_{\rm in}$ shrinks, the magnetic Purcell factor can be largely enhanced with a red shift of the resonance wavelength. At $r_{\rm in}=20\, {\rm nm}$, the peak Purcell factor can be larger than 200. While, for the electric Purcell effect [Fig. \ref{fig:Str}(f)], the decrease of $r_{\rm in}$ just slightly changes the peak Purcell factor and brings a blue shift of the resonance.


\subsection{\label{sec:level2}Insensitivity of Purcell factors to emitter's position}
In the following, we investigate the Purcell factors when the emitter moves along several paths in our nanostructure. It is found that the Purcell factor is high in a range of several-100-nm in the near-field region, which shows obvious changes of MLDOS and ELDOS. The deviation of emitter from center (of nanoring or gap) can even bring higher magnetic and electric Purcell enhancement. The results demonstrate the tolerance of emitter position in our hybrid nanostructure and thus provide more convenience in nano-fabrications. The large magnetic Purcell enhancement can be achieved inside nanoring and in the hollow region \cite{wu_strong_2019}. While, the electric Purcell enhancement keeps a high level when EQE locates around nanorod \cite{curto_unidirectional_2010}. These hotspots of enhancement also correspond to the field hotspots in magnetic and electric-like resonances [Fig. \ref{fig:Res}(b, d)]. 

\begin{figure}
	\center
	\includegraphics[width=0.85\textwidth]{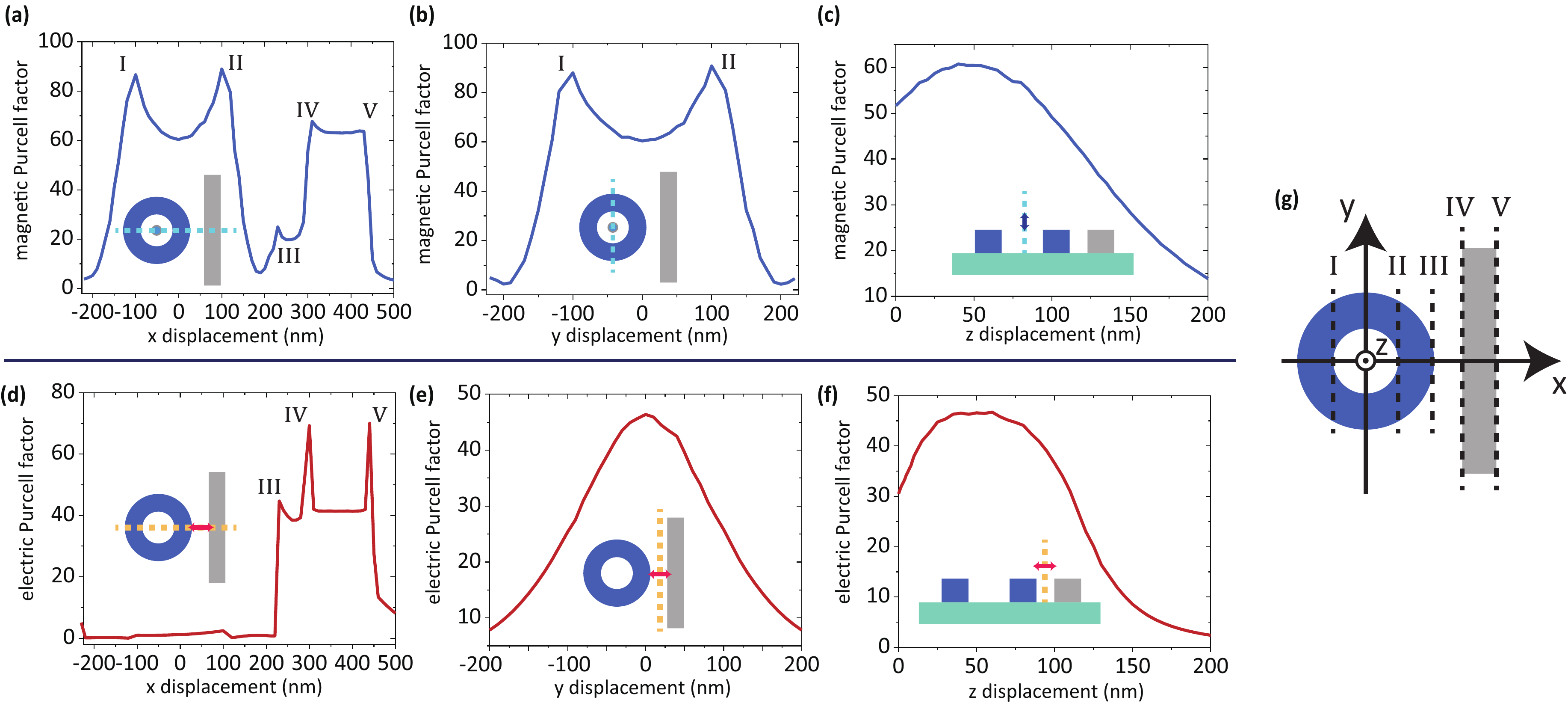}
	\caption{Variation of the magnetic and electric Purcell enhancement when the emitter’s position moves along several paths. (a-c) Magnetic Purcell factors when MQE moves along a diameter of nanoring normal to nanorod, parallel to nanorod or a path normal to substrate. In the situations above, MQE feels the strongest emission enhancement at the inner boundaries. The deviation from the center can increase the emission enhancement, thus the emitter's position has a large tolerance. (d-f) Electric Purcell factors when EQE moves along a nanoring diameter normal to nanorod, parallel to nanorod or a path normal to substrate. EQE feels no enhancement near the nanoring. The peak appears at the boundary of nanorod and EQE feels a stronger enhancement when it deviates from the middle of gap. (g) Illustration of some specific points: the origin is set at the center of nanoring, and the origin of normal direction is set at the surface of substrate. Some specific positions are also ticked out by Roman numberals: I/II: left/right inner boundary of nanoring, III: right outer boundary of nanoring, IV/V: left/right boundary of nanorod.}
	\label{fig:Tol}
\end{figure}

The variation of magnetic and electric Purcell factors is shown below. Consider the magnetic Purcell effect, the Purcell factors keep at a high level around nanoring, which can be verified in their variation along several typical paths. Firstly, we choose a path through center of nanoring and normal to nanorod [Fig. \ref{fig:Tol}(a)]. When MQE moves along the path, the Purcell factors keep increasing inside nanoring and it reaches maximums with Purcell factor around 90 at inner boundaries. It can be seen that the deviation from center does no harm to but benefits the higher emission enhancement. Then, when MQE locates in the gap region, the higher MLDOS occurs in the region close to nanorod. And the Purcell factors keep steady at 60 when MQE intrudes into nanorod. Secondly, when MQE moves along a diameter parallel to nanorod [Fig. \ref{fig:Tol}(b)], the peaks of MLDOS appear at inner boundaries of nanoring with Purcell factor around 90. Thirdly, when the distance to substrate of MQE changes, the transition of Purcell factor is shown in Fig. \ref{fig:Tol}(c). For the maximal Purcell enhancement, MQE can be placed 50 nm above substrate where the Purcell factor reaches 60. When MQE is closer to substrate or comes into substrate, the Purcell factor exhibits a decrease. 

The electric Purcell enhancement is also insensitive to the position of EQE, and keeps high when EQE lies around nanorod. As EQE moves along the path through center of nanoring and normal to nanorod, there is nearly no ELDOS enhancement when EQE is close to nanoring. Then the Purcell factor shows an increase when EQE locates in the vicinity of nanorod \cite{hu_enhancement_2015}. In our calculation, the electric Purcell factor approaches 70 at nanorod boundaries [Fig. \ref{fig:Tol}(c)]. Besides, where EQE moves through the gap [Fig. 3(d)], the highest Purcell factor reaches 45 at the gap center. For both the large Purcell enhancement and experimental practicability, EQE is suitable to locate in the gap. And in real experiments, it cannot be positioned too close to the nanorod boundary because it may induce quantum effects \cite{savage_revealing_2012, takeuchi_operation_2019} which is not included in the electromagnetic simulation. The influence of distance of EQE from substrate is demonstrated in Fig. \ref{fig:Tol}(f), where the trend of Purcell enhancement is similar with that of MQE. The largest enhancement appears when EQE locates at 40 nm above substrate. And when EQE moves down into substrate or away from structure, the Purcell enhancement experiences a drop.


\section{\label{sec:level1}Magnetic and electric Purcell enhancement and far-field radiation in an array}
\begin{figure}
	\center
	\includegraphics[width=0.85\textwidth]{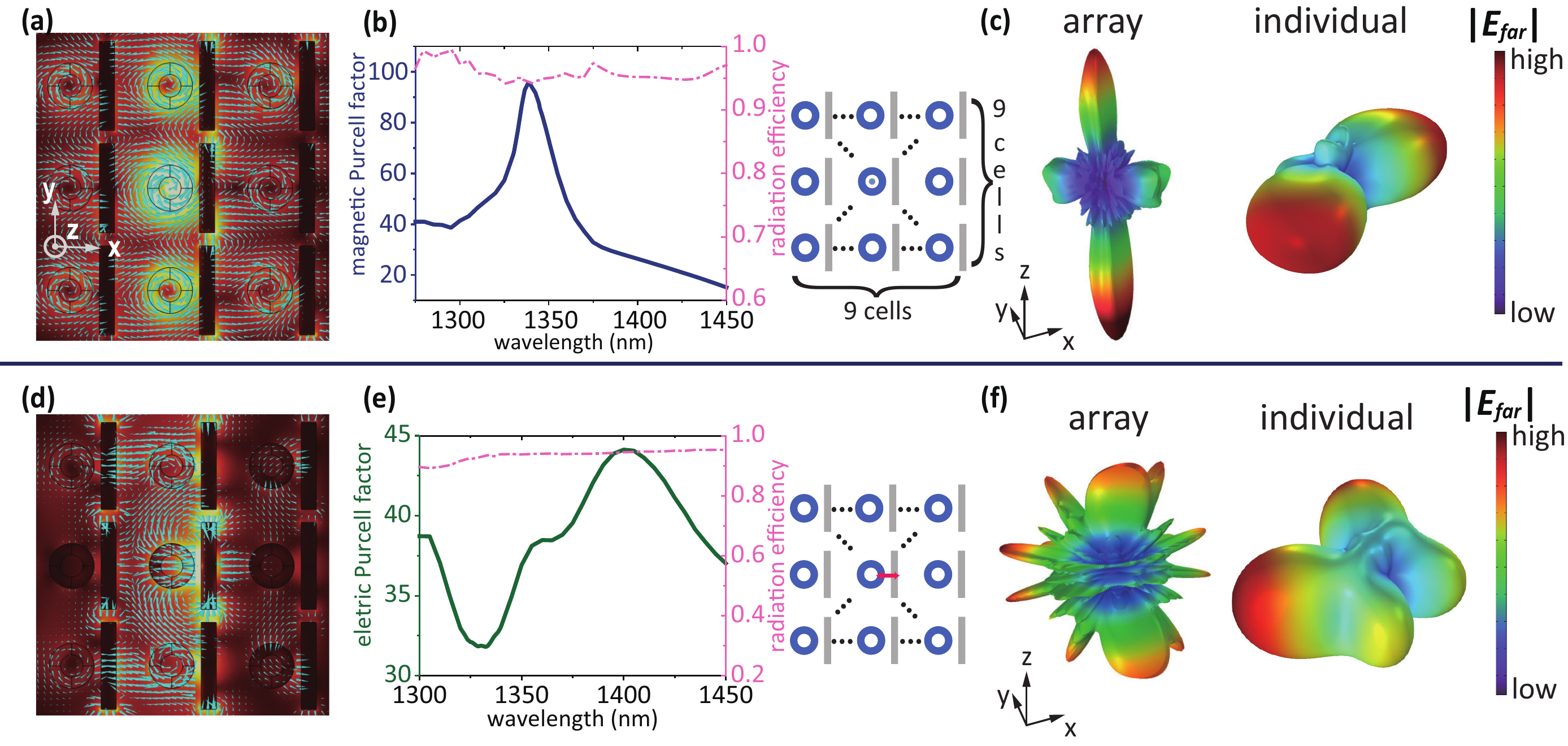}
	\caption{Purcell enhancement, radiation efficiency and far-field radiation in a $9\times9$ array of hybrid structures. (a, d) Field profiles of the array excited by MQE (or EQE) situated at the central cell. MQE mainly excites the magnetic-like resonances in adjacent cells. EQE excites the electric-like resonances at adjacent cells in the same column, but in other cells the excitation is relatively weak. (b, e) Radiation properties including the magnetic (electric) Purcell factors and radiation efficiency. The right insets give schematics of the $9\times 9$ array excited by MQE or EQE. Compared with an individual structure, the magnetic Purcell factor shows a larger enhancement and reaches a peak of 97. EQE can yield a large emission enhancement by 45 folds. The radiation efficiency stays above 90\% in the both setups. (c, f) Far-field radiation of the array (or the individual structure) excited by MQE or EQE. Compared with the individual structure, the far-field radiation exhibits a more directional radiation when the array is excited by MQE or EQE.}
	\label{fig:Array}
\end{figure}

In this part, we explore the emission properties in an array composed of the hybrid structures [Fig. \ref{fig:Array}(a)]. The emission of a single MQE or EQE can be modulated under the mode hybridization joined by cells in the array. The magnetic Purcell factor even has a further increase in the array. The far-field radiation from MQE in this array also possesses a high directivity.
Every cell in the array is an individual hybrid structure with the same structural parameters as above, and the intervals between rows and columns are 1000 nm. We simulate a $9\times9$ array with its size of $9650 \,\rm nm\times 9650 \,\rm nm\times 2100 \,\rm nm$ in the electromagnetic simulation. Only one MQE or EQE is set to excite the entire array, which locates in the central cell. Additionally, MQE lies at the center of nanoring, and EQE is at the middle of the gap.
The only MQE or EQE locates in the central cell [ticked out in the insets of Fig. \ref{fig:Array}(b, e)]. We choose $9\times9$ size to ensure that Purcell factor keeps at same level when the array size expands, and it is not oversized that avoids unnecessary calculations (more details in Appendices).

When the array is excited by MQE at the center of nanoring, it shows its advantages in the larger magnetic Purcell enhancement and effective far-field radiation. There mainly exist the magnetic-like resonance in one cell, which can be seen in Fig. \ref{fig:Array}(a). The maximal Purcell enhancement is close to 100 folds at 1330 nm, higher than the situation of an individual structure mentioned above. The array provides a stronger Purcell enhancement because of the higher MLDOS enhancement with the existence of more adjacent structures \cite{du_optical_2021}. While, for the far-field radiation, the radiation efficiency maintains above 90\%. The radiation [Fig. \ref{fig:Array}(c)] exhibits a good directional pattern to normal direction with side lobes only distributed at in-plane directions, where about 70\% of the far-field emission radiates into normal direction. Compared with the individual structure, the array can emit more directional radiation by influence of more adjacent cells.

When EQE in the gap excites the array, the maximal Purcell factor keeps the same level compared with an individual structure, which reaches 45 around 1400 nm. Excited by EQE, the electric-like resonance dominates in the central cell. The ELDOS enhancement is not largely enhanced by the array because the excitation is relatively weak in adjacent and remoter cells [Fig. \ref{fig:Array}(d)], which stems from high absorption induced by metallic rods. The radiation efficiency also stays above 89\% indicating that the metal absorption is also suppressed in the array. As shown in Fig. \ref{fig:Array}(f), the radiation pattern also exhibits the increase of directivity compared with the situation in the individual structure \cite{vaskin_light-emitting_2019}. 

\section{\label{sec:level1}Conclusion}
Here, we propose a hybrid metal-dielectric structure composed of a silicon nanoring and a metal nanorod. The structure supports both large MLDOS and ELDOS, which can produce simultaneously 60-fold magnetic Purcell enhancement and 45-fold electric Purcell enhancement with a 70-nm gap. Its advantages show in the stronger field confinement and relatively low Ohmic loss compared to the situation of a single dielectric nanoring or metal nanorod. Moreover, the array composed of hybrid structures can modify the emission properties in directional far-field radiation and larger Purcell enhancement. 

The structure is possible by current nanofabrication techniques such as electron beam lithography \cite{yang_all-dielectric_2014, ho_highly_2018}. The emitter can be prepared by doping nanocrystals and it can be assembled into the nanostructure using atom force microscopy tips \cite{aigouy_mapping_2014}. The hybrid structure also bring conveniences to nano-fabrication. The limit of positioning of emitters can be relaxed because of the high MLDOS and ELDOS in a wide region around structure. The arrangement is relatively simple and also provides rich parameters to modulate resonances for optical antennas and metasurfaces. Our proposed hybrid metal-dielectric structure utilizes the hybridization of MD and ED, EQ resonances in dielectric and metal structures, which enriches options to enhance magnetic emission and modify radiation patterns at nanoscale. The magnetic and electric Purcell enhancement also provides a route to selective excitation for photon sources on integrated circuits. It may also facilitate the development of brighter magnetic emission sources and light-emitting metasurfaces.

\begin{acknowledgments}
This work was supported by the National Natural Science Foundation of China (11974032, 11734001, 11525414); Key R\&D Program of Guangdong Province (2018B030329001).
\end{acknowledgments}

\appendix

\setcounter{figure}{0}
\renewcommand{\thefigure}{a\arabic{figure}}

\section{Resonances of a silicon nanoring or silver nanorod}
\begin{figure}
	\center
	\includegraphics[width=0.7\textwidth]{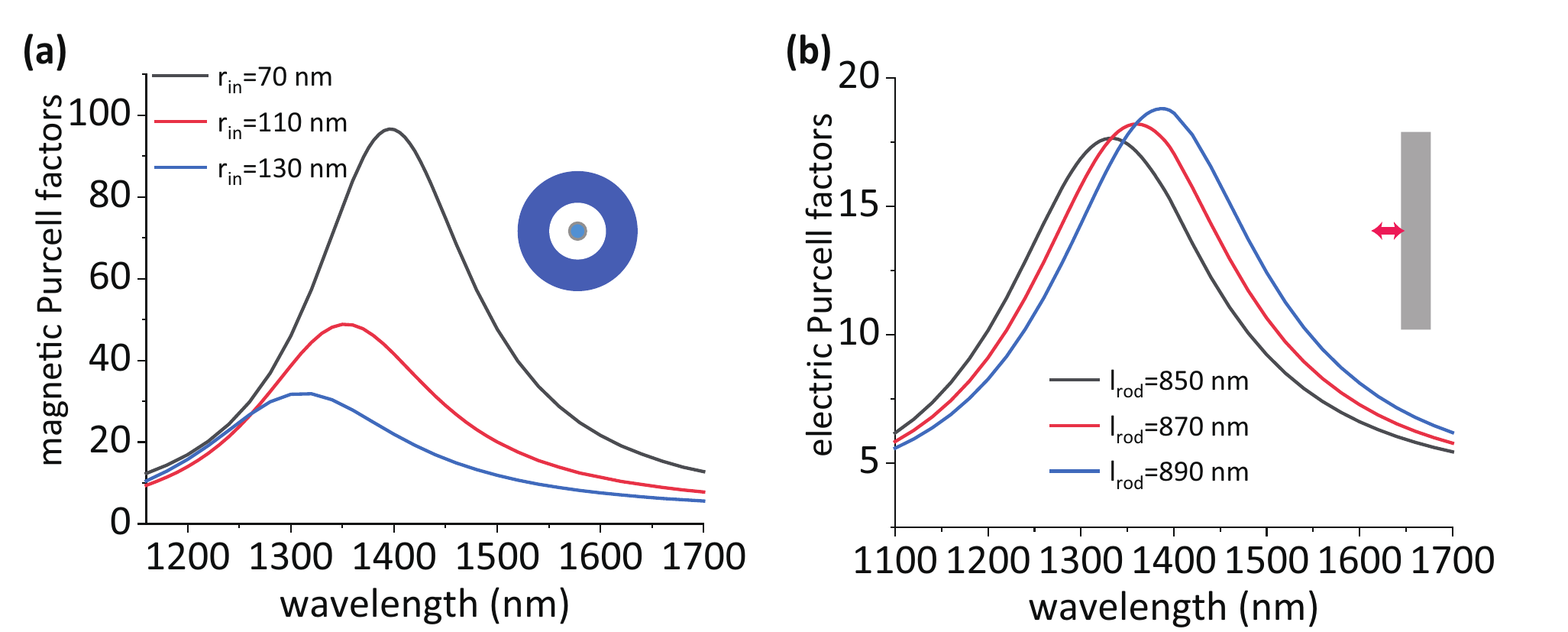}
	\caption{Left(Right): Variations of magnetic(electric) Purcell factors when MD(EQ) resonance of the silicon nanoring(metal nanorod) excited by MQE(EQE). MD resonance of nanoring shows a blue shift when the inner radius of nanoring $r_{\rm in}$ enlarges, and EQ resonance of nanorod shows a red shift when the rod length $l_{\rm rod}$ enlarges.}
	\label{fig:a1}
\end{figure}

For a single silicon nanoring or silver nanorod, the geometrical parameters are chosen that MD of nanoring and EQ of nanorod both locates at 1360 nm (red solid lines in Fig. \ref{fig:a1}). When the two components are involved into mode hybridization, the derived magnetic-like and electric-like resonances are situated in 1305 nm and 1430 nm, respectively. Fig. \ref{fig:a1} depicts Purcell factors when MD of silicon nanoring or EQ of silver nanorod is excited by MQE or EQE. When the inner radius of nanoring enlarges, MD resonance shows a blue shift. And EQ resonance shows a red shift when the rod length $l_{\rm rod}$ enlarges. These results correspond to the previous research \cite{feng_all-dielectric_2016, curto_unidirectional_2010} and validates our numerical simulation.

\section{Influence of Purcell enhancement by the array size}
The size of array composed of the hybrid structures will influence the resonance of entire array and the Purcell enhancement of MQE and EQE \cite{vaskin_light-emitting_2019}. We check the $6\times 6, 9\times 9$ and $12 \times 12$ arrays, and the Purcell enhancements keep at the level higher than those in individual structures. Here, we choose the $9\times 9$ array in numerical simulations by virtue of computing resources. The resonance of the array is influenced by the enlargement of array size. 

\begin{figure}
	\center
	\includegraphics[width=0.7\textwidth]{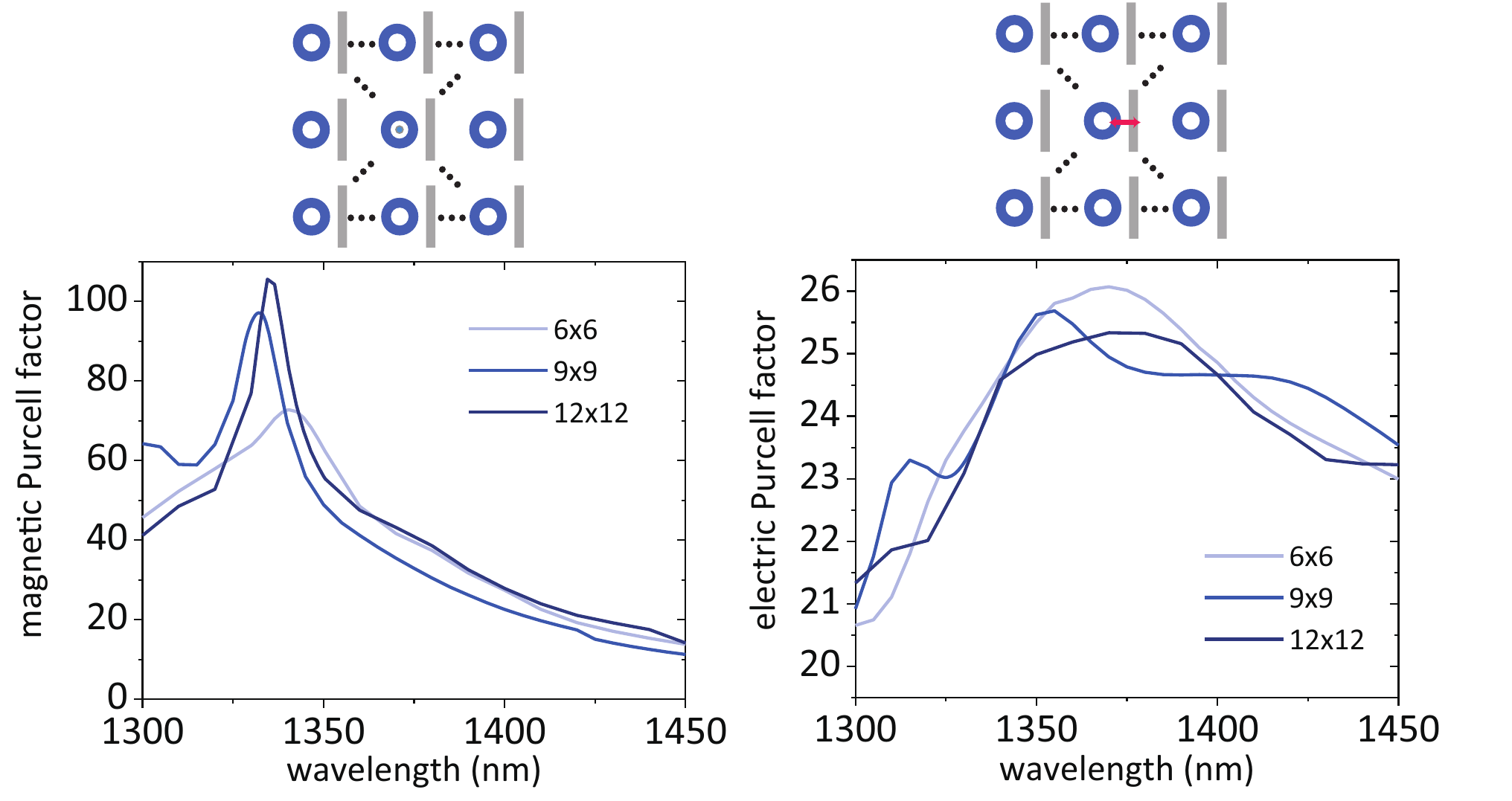}
	\caption{Left(Right): Variation of magnetic(electric) Purcell enhancement in a $6\times6$, $9\times9$ and $12\times12$ array.}
	\label{fig:a2}
\end{figure}





\bibliography{ref_hybridPurcell}

\end{document}